\documentclass[aps,prb,amsmath,amssymb,reprint,superscriptaddress]{revtex4-1}
\usepackage{graphicx}
\usepackage{dcolumn}
\usepackage{bm}
\usepackage[english]{babel}
\usepackage{color}
\usepackage{verbatim}
\renewcommand{\vec}[1]{\boldsymbol{#1}}

\begin{document}

\title{Graphene Electrodynamics in the presence of the Extrinsic Spin Hall Effect}

\author{Chunli Huang}
\affiliation{
 Division of Physics and Applied Physics, School of Physical and Mathematical Sciences,
Nanyang Technological University, Singapore 637371, Singapore
}
\affiliation{Department of Physics, National Tsing Hua University, Hsinchu 30013, Taiwan}

\author{Y. D. Chong}
\affiliation{
 Division of Physics and Applied Physics, School of Physical and Mathematical Sciences,
Nanyang Technological University, Singapore 637371, Singapore
}

\author{Giovanni Vignale}
\affiliation{
 Department of Physics and Astronomy, University of Missouri, Columbia, Missouri 65211, USA
}
\author{Miguel A. Cazalilla}
\affiliation{Department of Physics, National Tsing Hua University and National Center for Theoretical Sciences, Hsinchu 30013, Taiwan}

\date{\today}

\begin{abstract}
We extend  the electrodynamics of two dimensional electron gases to account for the extrinsic spin Hall effect (SHE). The theory is applied to doped graphene
decorated with a random distribution of absorbates that induce spin-orbit coupling (SOC) by proximity.  The formalism  extends previous semiclassical treatments of  the SHE to the non-local dynamical regime. Within a particle-number conserving approximation, we compute the conductivity, dielectric function, and spin Hall angle  in the small frequency and wave vector limit. The spin Hall angle is found to decrease with frequency and wave number, but it remains comparable to 
its zero-frequency value around the frequency corresponding to the Drude peak. 
The plasmon dispersion and linewidth are also obtained.  The extrinsic SHE affects the plasmon dispersion in the long wavelength limit, but not at large values of the wave number. This result suggests an explanation for the rather similar plasmonic response measured in exfoliated graphene, which does not exhibit the SHE, and graphene grown by chemical vapor deposition, for which a large SHE has been recently reported. Our theory also lays the foundation for future experimental searches of SOC effects in the electrodynamic response of two-dimensional electron gases with SOC disorder.
 \end{abstract}
\pacs{}
\maketitle

\section{Introduction}

A number of ground-breaking experiments have recently explored the physics of plasmons in doped Dirac materials like graphene and topological insulators.\cite{stauber2014plasmonics,ju2011graphene,di2013observation,grigorenko2012graphene,chen2012optical}  One major difference between these two kinds of Dirac materials is the strength of the spin-orbit coupling (SOC) in their band structure. Whilst SOC is substantial in the band structure of topological insulators, it  has been rightly dismissed in graphene as negligible due to the lightness of carbon.~\cite{huertas2006spin,min2006intrinsic}

  Nevertheless, despite of the negligible SOC in its band structure, it has been recently observed that  the spin Hall effect (SHE) can occur in graphene decorated with absorbates such as hydrogen,~\cite{balakrishnan2013colossal} and in graphene devices prepared by chemical vapor deposition (CVD),~\cite{jaya2014giant} which contain residual metallic clusters. From the theoretical point of view, 
the existence of a (proximity-induced) SOC in graphene is also expected from first-principles and tight-binding  calculations.~\cite{PhysRevX.1.021001,brey2015,PhysRevLett.109.116803} However, the experimental situation remains rather controversial,~\cite{fuhrer2015,vanwees2015} and in this context,
deeper theoretical studies of decorated graphene are, in our opinion, more than necessary.

 A direct experimental consequence of the SOC disorder is the SHE by which an applied electric field induces a transverse spin current.~\cite{sinova2015review,dyakonov1971current,hirsch1999spin,zhang2000spin,vignale2010ten}  The figure of merit of the SHE is the spin Hall angle, $\Theta_\mathrm{SH}$ which measures the fraction of the (longitudinal) charge current that is converted into a (transverse) spin current. In Ref.~\onlinecite{jaya2014giant}, a large value of $\Theta_\mathrm{SH} \sim 0.1$ was reported, which makes CVD graphene a material comparable to platinum, a reference material in the field of metallic spintronics.~\cite{sinova2015review} 
In this field, over the last decade, the SHE has attracted increasing theoretical \cite{schliemann2006spin,PhysRevLett.103.026804,Ferreira2014} and experimental \cite{kato2004observation,wunderlich2005experimental,seki2008giant,jaya2014giant,balakrishnan2013colossal} attention, and the most recent experimental discoveries may point the way toward a new class of graphene-based spintronic devices.\cite{liu2011spin,liu2012spin}

This paper examines corrections to electrodynamics arising from extrinsic SHE. The latter is known to occur in materials through essentially two kinds of  microscopic mechanisms, namely side-jump and skew scattering.\cite{RevModPhys.82.1539} In the doped regime, the large SHE observed in adatom decorated and CVD  graphene \cite{jaya2014giant,balakrishnan2013colossal,PhysRevLett.109.116803,PhysRevLett.110.246602,PhysRevB.91.115141} can be attributed mainly to skew-scattering, since the total adatom coverage  is typically very dilute and the side-jump contribution to the SHE is subleading in this regime.~\cite{jaya2014giant,sinova2008,sinova2015review} The skew-scattering is strongly enhanced by the occurrence, in the neighborhood of the Dirac point, of scattering resonances caused by adatoms and other impurities.~\cite{Ferreira2014,jaya2014giant} 

In earlier work that analyzed the electrodynamic response of doped graphene by focusing on the effects of disorder, skew-scattering effects 
have been neglected, to the best of our knowledge. For instance,  the recent analysis by Principi~\emph{et al.} examined the effects of disorder on graphene plasmons,~\cite{principi2013impact} but did not take into account the losses  and other effects that may arise from SOC disorder. From a theoretical point of view, it can argued that a description neglecting the extrinsic SHE  is substantially incomplete. This is because a sizable fraction of the oscillatory electric current generated by the plasmon is converted into a spin current by the SHE.  Such a scenario is plausible because, in the long wavelength limit, the plasmon frequency in a two dimensional electron gas reaches arbitrarily low frequencies (i.e. the plasmon dispersion is $\omega_p(q) \sim \sqrt{q}$, where $q$ is the wave number). Therefore, we can approximately rely on the DC value of the spin Hall angle to estimate the charge to spin current conversion. On the other hand, it can be also argued that spin currents are charge neutral and, as a consequence, they should not modify the electrodynamic response of the system. Below we shall see that, although the latter expectation is sensible, the physics is more subtle and electrodynamics is indeed modified by the SHE. 

 In addition, beyond the low frequency regime, where the stationary (DC) theory of the SHE may be approximately valid, a theory capable of accounting for the full frequency dependence of the spin Hall angle needs to be formulated. Previous theoretical treaments of extrinsic SHE in graphene~\cite{Ferreira2014,jaya2014giant,hungyu2015} have focused on DC transport properties. However, the AC regime remains unexplored. Such a task is undertaken in this work, focusing on the case of doped graphene. With minor modifications, our theory can also be applied to other types of two-dimensional electron or hole gases with SOC disorder. Therefore, studies like the present one, may one day allow for spin-current generation in decorated graphene by means of plasmon excitation, a possibility which has been recently considered in the spintronics literature.~\cite{uchida2015generation}

Similar to Refs.~\onlinecite{Ferreira2014,hungyu2015}, we consider a single layer of doped graphene decorated with (clusters of) absorbates which induce SOC by proximity.  Our analysis relies on the semi-classical Boltzmann transport equation (BTE) in the dilute-impurity limit. Using a general form of the single-impurity $T$-matrix,~\cite{Ferreira2014,hungyu2015} we solve the  time-dependent linearized BTE within a particle-conserving approximation.~\cite{MerminPhysRevB.1.2362} We find that in the presence of SOC disorder, the collision integral of the BTE contains two different contributions: (i) a conventional momentum (Drude) relaxation term, and (ii) an effective Lorentz force arising from the skew scattering with SOC disorder. From the solution of the BTE, we derive a generalized Ohm's law in the AC regime, as well as the dynamic nonlocal charge conductivity, dielectric function, and the AC spin Hall angle. 

Our theory applies to the THz regime, which corresponds to the typical plasma frequency for (moderately) doped graphene.\cite{ju2011graphene} The excitations in this frequency regime are dominated by the contribution of the $\pi$-band electrons, which are strongly hybridized with the outer shell electrons of the absorbates. As a result of this hybridization, the SOC disorder is induced. Therefore, our present theory does not apply to visible or higher frequencies for which electronic excitations localized in the absorbates that decorate graphene can be important. Furthermore, our theory does not apply to devices in which metallic nano-particles are deposited on top of graphene covered by a dielectric layer. In such devices, the coupling between the nanoparticles and graphene is essentially capacitive (as the insulating layer strongly or completely suppresses the hybridization of the nanoparticles with the graphene $\pi$-band), and therefore there is no proximity-induced SOC.

In order to obtain the plasmon dispersion and linewidth, we numerically find the  zeros  of the dielectric function in the complex frequency plane. Generally speaking, disorder has two major known effects on the plasmon properties.  First, it contributes to the linewidth (i.e. imaginary part of the plasmon frequency),  making the plasmon lossier.~\cite{principi2013impact} Second, it ``softens'' the plasmon dispersion by reducing the plasma frequency $\omega_p(q)$ relative to the clean system  for each value of  $q$. Thus, in the presence of disorder, the  plasmon frequency vanishes at a cutoff wave number $q_{c}$. Similar softening has been previously predicted in disordered two-dimensional (2D) electron gases\cite{quinn_giuliani} and spin-polarized graphene.~\cite{giovanni_PhysRevB.90.155409} 
We find that SOC disorder reduces the plasmon linewidth 
and it modifies the plasmon softening by reducing the softening at low wave numbers and increasing it at large wave numbers, relative to the case of purely scalar disorder. These effects are relatively modest; the shifts in plasmon frequency, for instance, are of order 1\% at experimentally realistic frequencies and SOC disorder strengths.
We believe that  this may explain why exfoliated~\cite{fei2012gate} and CVD graphene~\cite{chen2012optical} have been experimentally found to exhibit a similar plasmonic response, despite the rather large SHE effect recently observed in CVD graphene.~\cite{jaya2014giant}

The rest of the article is organized as follows. In Sec.~\ref{sec:results}, we summarize  the most important results of the article, namely the AC generalization of the Ohm's law in the presence of SOC disorder. The effect of SOC disorder on the plasmon dispersion and lifetime is also discussed in Sec.~\ref{sec:results}, together with the long wavelength limit of the conductivity, dielectric function, and spin Hall angle. In Sec.~\ref{sec:BTE}, we provide the details of the derivation of the results described in Sec.~\ref{sec:results}.  We close the article with a summary and an outlook in 
Sec.~\ref{sec:summ}. The most technical parts of these derivations have been relegated to the Appendices. Throughout, we work in units where $\hbar=1$.

\section{Results}
\label{sec:results}

In this section, we summarize the main results of this work and discuss some of their experimental consequences.  The derivations for the equations presented here are given in the following sections, whist the most technical details have been relegated to the Appendices. We begin by introducing a generalized Ohm's law which relates the charge and spin currents. Hence, the form of AC charge conductivity, dielectric function, and spin Hall angle follow.  The modified plasmon dispersion and linewidth, as well as the conductivity in the optical limit are also discussed 
in this section.

\subsection{Generalized Ohm's law}


In graphene with extrinsic SOC, we find the following generalized AC Ohm's law:
\begin{align} 
\vec{J}_{c}(q,\omega) &= \sigma_m(q,\omega) \left[\vec{E}(q,\omega) - \rho_{ss} \vec{J}_{s}(q,\omega) \times \vec{\hat{z}} \right],\label{eq:result1}\\
\vec{J}_{s}(q,\omega) &=\Theta_{\mathrm{SH}}(q,\omega) \: \vec{J}_{c}(q,\omega) \times  \vec{\hat{z}}  .
 \label{eq:result2}
\end{align}
Here $\vec{E}(q,\omega)$ is the electric field, $\vec{J}_{c}(q,\omega)$ is the charge current density, $ \vec{J}_{s}(q,\omega )$ is the transverse spin current density, and $q$ and $\omega$ are the wave-number and frequency respectively; $\sigma_{m}(q,\omega)$ is the contribution to the charge conductivity from scalar disorder, $\Theta_{\mathrm{SH}}(q,\omega)$ is a dynamic spin Hall angle, and $\rho_{ss}$ is the skew-scattering resistivity, which is a constant. Explicit expressions for $\sigma_{m}$, $\Theta_{\mathrm{SH}}$, and $\rho_{ss}$ are given in Sec.~\ref{sec:BTE}.  For the moment, we simply note that they depend parametrically on  the elastic mean free (Drude) time,  $\tau_\parallel$ , and the zero temperature  DC spin Hall angle:
\begin{equation}
  \gamma = \frac{\tau_\parallel}{\tau_\perp} = \Theta_{\mathrm{SH}}(q=0,\omega=0),
\end{equation}
where $\tau_\perp$ is the skew-scattering time; the relationship of $\tau_{\parallel}$ and $\tau_{\perp}$ to the properties of the disorder is explained in Appendix \ref{app:lifetime}. The parameter $\gamma$ increases with the strength of the SOC disorder.  In the limit  $\gamma\to 0$, both $\rho_{ss}$ and $\Theta_{\mathrm{SH}}(q,\omega)$ vanish, and $\sigma_{m}(q,\omega)$  becomes the charge conductivity $\sigma_{c}(q,\omega)$.

\begin{figure}
\includegraphics[width=8cm]{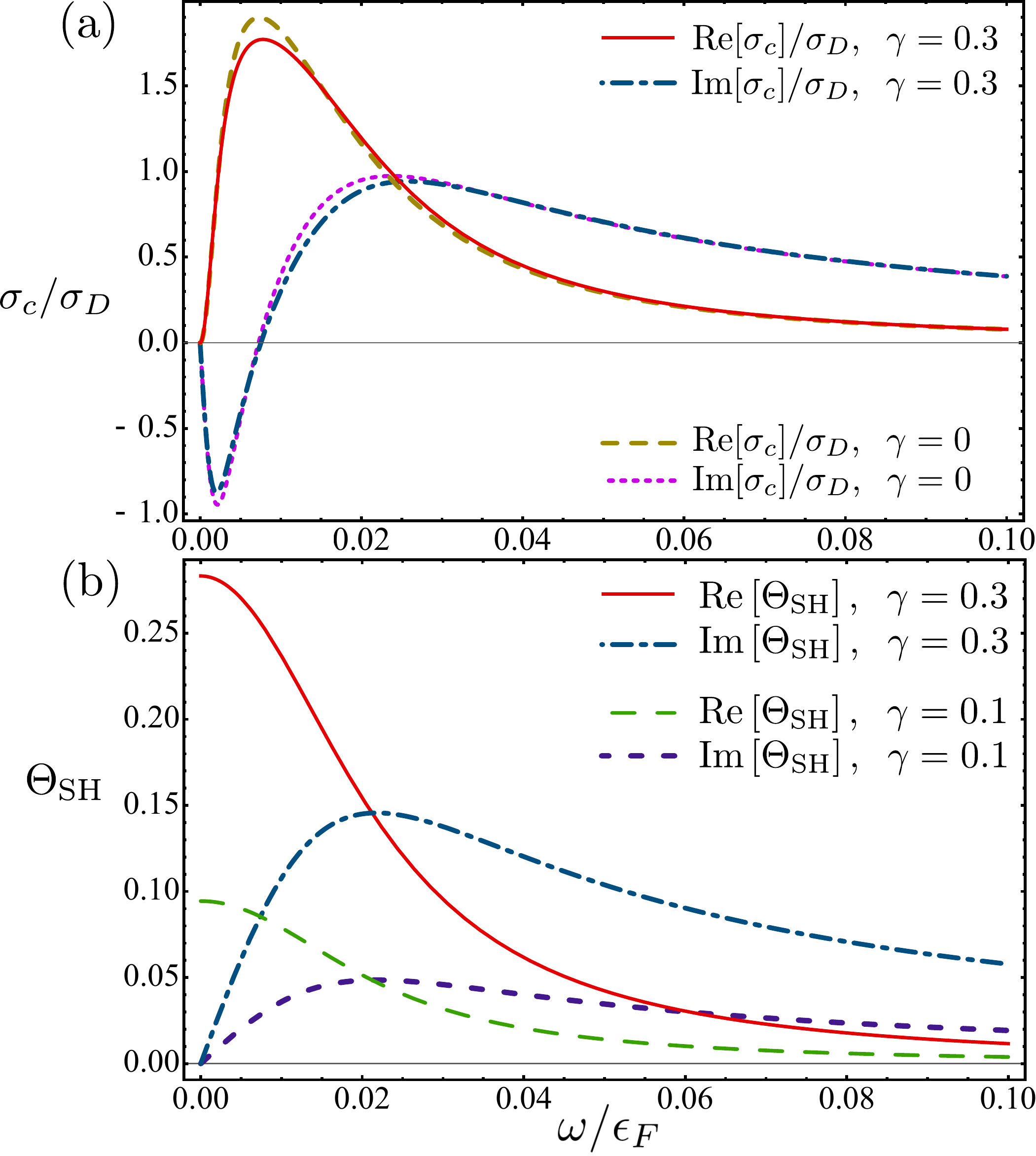}
\caption{\label{fig:die_func}
  (a) Frequency dependence of the dynamic charge conductivity $\sigma_{c}$, normalized to the Drude conductivity $\sigma_{D}$.  (b) Frequency dependence of the dynamic spin Hall angle $\Theta_{\mathrm{SH}}$.  Note that for frequencies larger than those  corresponding to the Drude peak in the conductivity, the real part of 
 $\Theta_{\mathrm{SH}}$ decays much faster than the imaginary part. In this plot $q=0.01 \: k_F$ and the elastic collision rate $\tau_\parallel = 5\: \epsilon_{F}^{-1}$, where $\epsilon_F$  ($k_F$) is the Fermi energy (momentum). The sign change of the imaginary part of the conductivity for $q\approx 0.01 \: k_F$ is due to the crossover from the collisionless  to the hydrodynamic regime.}
\end{figure}

We can substitute Eq.~\eqref{eq:result2} into \eqref{eq:result1} to eliminate the spin current $\vec{J}_s$.  This gives $\vec{J}_c (q,\omega)= \sigma_c(q,\omega) \vec{E}(q,\omega)$, where
\begin{equation}
\sigma_{c}(q,\omega)=\frac{\sigma_{m}(q,\omega)}{1+ \rho_{ss}  \Theta_{\mathrm{SH}}(q,\omega) \sigma_{m}(q,\omega)}
\label{eq:sigmac}
\end{equation}
is the dynamic charge conductivity. The dielectric function can be obtained using the continuity equation, $-i\omega \rho_c (q,\omega)+ i \vec{q}\cdot \vec{J_c} (q\omega)= 0$, where $\rho_c(q,\omega)$ is the charge density. This yields:
\begin{equation}
\epsilon(q,\omega)=1-\frac{\omega_{p}^{2}(q)}{v_{F}^{2}q^{2}}\left( \frac{\chi_{m}(q,\omega) / N(\epsilon_{F})}
{1+ \rho_{ss} \sigma_{m}(q,\omega)  \Theta_{\mathrm{SH}}(q,\omega) }\right),\label{eq:dielec}
\end{equation}
where $\omega_{p}(q)=\sqrt{4\alpha_{gr}k_F v_{F}^{2} q}$ is the plasma frequency of pristine graphene; $\alpha_{gr}=e^{2}/v_{F}$, $N(\epsilon_{F})=2\epsilon_{F}/\pi v_{F}^{2}$ is the density of states at the Fermi energy, and $\chi_{m}(q,\omega) = q^2 \sigma_m(q,\omega)/i e^2 \omega$ is the Lindhard function including the correction from the 
particle-number conserving relaxation-time 
approximation~\cite{MerminPhysRevB.1.2362} (see Sec.~\ref{sec:generalized_ac_formula}). Note that for $\gamma =0$, equations~\eqref{eq:sigmac} and \eqref{eq:dielec} reduce to the known results for a 2D electron gas with scalar disorder
(see e.g. Ref.~\onlinecite{giuliani_giovanni}). 

A typical plot of $\sigma_c(q,\omega)$ as a function of $\omega$, for finite $q$, is shown in Fig.~\ref{fig:die_func}(a). The real part of the conductivity exhibits a Drude peak around the plasma frequency and 
the imaginary part of the conductivity changes sign around the Drude peak. At low frequencies, the system is in the hydrodynamic regime and the imaginary part of the conductivity is negative. However, at high frequencies, the system is in the collisionless  regime and the imaginary part of the conductivity is positive. Thus, the sign change in the imaginary part of the conductivity around the Drude peak corresponds to the crossover from the diffusive to collisionless regime. The plot also shows that the corrections to the conductivity  due to the extrinsic SHE are small, and most pronounced around the
Drude peak frequency. 

\begin{figure}
\includegraphics[width=8.5cm]{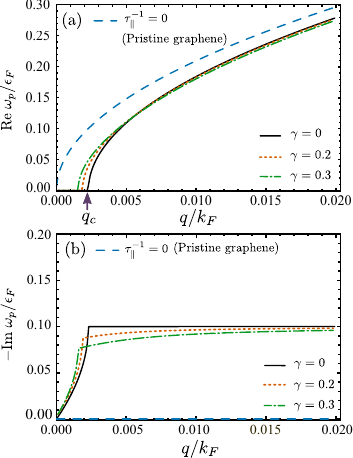}
\caption{\label{fig:plasmon_dis}
  Plasmon dispersion relation (a) and linewidth (b) in doped graphene with SOC disorder leading to an extrinsic SHE, for three different values of $\gamma = \tau_{\parallel}/\tau_{\perp}$,  which measures ratio of charge current converted into spin current in the DC limit at zero temperature. The results are plotted for a value of elastic mean free time $\tau_{\parallel}= 5 \: \epsilon^{-1}_F$. As shown in the top panel, increasing $\gamma$ does not have a dramatic effect on the plasmon dispersion and linewidth except near the cutoff wave number $q_c$. The black curve is the plasmon dispersion for pristine graphene. 
}
\end{figure}

Returning to Eq.~\eqref{eq:result1} and Eq.~\eqref{eq:result2}, they can be rearranged as follows:
\begin{equation}
\frac{\vec{J}_{c}(q,\omega)}{\sigma_{m}(q,\omega)} + \rho_{ss} \vec{J}_{s}(q,\omega) \times \vec{\hat{z}} =  \vec{E}(q,\omega),
\end{equation}
which makes it manifest that the response to AC electric field involves, not only a longitudinal AC charge current, but also a transverse AC spin current, as a result of skew scattering.  This is the  generalization of the SHE  to the AC regime. In particular, notice that the conversion efficiency is determined by the dynamic spin Hall angle $\Theta_{\mathrm{SH}}(q,\omega)$, plotted in 
Fig.~\ref{fig:die_func}(b). It can be seen that the dynamic spin Hall angle diminishes with increasing $\omega$, but remains comparable in magnitude to its DC value around the frequency of the Drude peak. In the DC SHE, the spin current generated by the SHE accumulates at the edge of the sample, and in two dimensional electron gases it has been detected by measurements of Kerr rotation \cite{kerr_rot_Choi} or non-local conductance.\cite{jaya2014giant}  The AC effect is more challenging to detect, since an oscillating spin current does not produce spin accumulation, but an oscillating magnetization instead.  One possible experimental signature, which we  discuss below, may be the modification of the plasmon dispersion relation due to the extrinsic SHE.

\subsection{Plasmon Frequency and linewidth}
\label{sec:Plasmon Frequency and Line-width}

We  can numerically calculate the plasmon frequency and linewidth  from the complex zeros of the dielectric function, Eq.~(\ref{eq:dielec}). The results  are displayed in Fig.~\ref{fig:plasmon_dis}.  The  effects of both scalar and SOC disorder on plasmon dispersion are to add a negative imaginary part to $\omega$ (accounting for a damping mechanism), and to ``soften'' (i.e. red shift) the plasmon frequency at a given wave number $q$ relative to its value in the clean system (cf. Fig.~\ref{fig:plasmon_dis}).  

In the absence of SHE (i.e. $\gamma = 0$), our numerical results agree with
the formula:~\cite{quinn_giuliani}
\begin{equation} \label{eq:plas_dis}
\omega_p(q) =\frac{1+\tilde{q}}{2+ \tilde{q}}\left(  \sqrt{4\alpha_{gr}\tilde{q}(2+\tilde{q})-\frac{1}{\tau_{\parallel}^2}} - \frac{i}{\tau_{\parallel}} \right).
\end{equation}
Here $\tilde{q}=q/4\alpha_{gr}k_F$ is the dimensionless ratio of the plasmon wave 
number, $q$, to the 
the Thomas-Fermi screening wave number, 
$\alpha_{gr}=e^{2}/v_{F}$ ($v_F$ being the Fermi velocity in graphene),  and 
$\tau_{\parallel}$ is the elastic mean free time. Eq.~\eqref{eq:plas_dis} was first obtained by Quinn and Giuliani~\cite{quinn_giuliani} for a 2D electron gas with scalar disorder. More recently, a similar plasmon softening has been found for spin-polarized graphene in 
Ref.~\onlinecite{giovanni_PhysRevB.90.155409}.

As shown in Fig.~\ref{fig:plasmon_dis}, the real part of the plasmon frequency, $\mathrm{Re} \: \omega_p \rightarrow 0$ vanishes at a finite (cutoff) wave-number $q_c$. Below $q_c$, the zero of the dielectric function associated with the plasmon becomes purely imaginary, i.e. the plasmon becomes completely overdamped. For this reason,  the cutoff wave number $q_c$ is not easily observable since the plasmon becomes a rather ill-defined collective excitation (i.e. $\mathrm{Im}\,\omega_p \gg \mathrm{Re} \, \omega_p$) before it reaches this wave number. 

However, when comparing the $\gamma = 0$ and $\gamma \neq 0$ cases, we notice that the overall effects of skew scattering on the plasmon dispersion and lifetime are strongest in the long wave-length limit (cf. Fig.~\ref{fig:plasmon_dis}). 
We find that skew scattering 
causes the softening of the plasmon to occur at smaller
wave vector than in the absence of any SOC disorder.
It is also interesting to observe in Fig.~\ref{fig:plasmon_dis}(b) that, as the wave number $q$  increases and consequently the plasmon frequency 
becomes larger, the shift in the plasmon dispersion
relative to the $\gamma =  0$  case changes sign for $q \simeq 0.006 \: k_F$. This happens (roughly) for the value where the plasmon
frequency $\omega_p(q)\tau_{\parallel} \approx 1$, which 
corresponds to the crossover from the hydrodynamic to 
collisionless regime. Skew scattering also modifies the
the plasmon linewidth as it can be seen in Fig.~\ref{fig:plasmon_dis}(b), by decreasing the
value of the losses relative to the case with scalar
disorder only (i.e. $\gamma = 0$).

\begin{figure}
  \includegraphics[width=8cm]{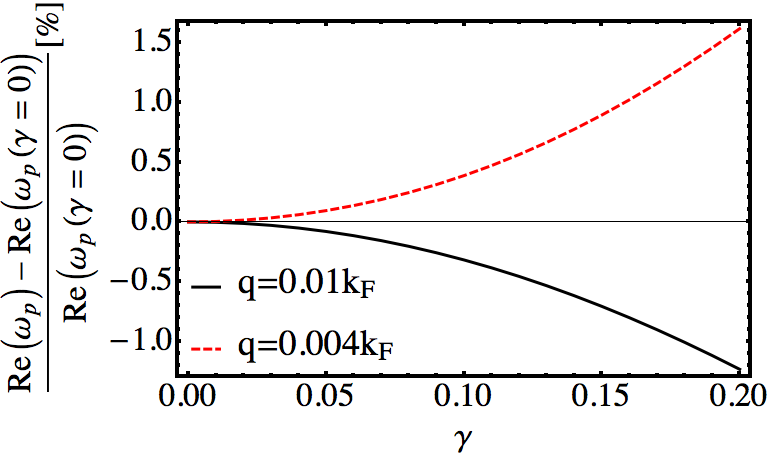}
  \caption{\label{fig:plasmon_soft}
Shift  in the plasmon frequency  relative to its value for $\gamma = 0$ (i.e. in the  absence of SOC disorder)  for $q=0.1k_{F}$ and $q=0.01k_{F}$. The effect of SOC disorder on the plasmon frequency is more pronounced in the hydrodynamic regime (where $\omega, v_{F}q\ll \tau^{-1}_{\parallel}$). The elastic mean free time is set to $\tau_{\parallel}= 5 \: \epsilon^{-1}_F$.}
\end{figure}

In order to better discern the effects of  SOC disorder from those of scalar disorder, in Fig.~\ref{fig:plasmon_soft} we have plotted the relative shift in the plasmon frequency, $\mathrm{Re}\: [\omega_{p} - \omega_p(\gamma=0)] / \mathrm{Re}\: \omega_p(\gamma=0)$ as a function of 
 $\gamma =\tau_{\parallel}/\tau_{\perp}$. The effect of SOC disorder on the plasmon frequency is more pronounced in the hydrodynamic regime (i.e. for $ v_{F}q \tau_{\parallel} \ll 1 $ and $\omega \tau_{\parallel} \ll 1$). However, for $ \gamma \approx 0.2$, which is comparable to the largest values reported in  Ref.~\onlinecite{jaya2014giant} for CVD graphene, the change in the plasmon frequency at $q = 0.01\: k_F$ is $\approx 1 \: \%$.  This small correction may account for the rather similar plasmonic response exhibited CVD and exfoliated graphene. We recall that for the latter, no appreciable SHE has been  observed to date (see, e.g. Ref.~\onlinecite{jaya2014giant}). Note as well that the corrections for $ q = 0.004 \: k_F$ and $q = 0.01\:k_F$ have  different signs, reflecting the crossover from the hydrodynamic to the collisionless regime taking place for $q\approx 0.006\: k_F$, as discussed above. 

\subsection{Response in long wavelength limit ($q \to 0$)}

The plasmon properties discussed in Sec.~\ref{sec:Plasmon Frequency and Line-width} are derived for an infinite graphene sheet in a spatially-uniform dielectric environment. However, in practical applications, other geometries, such as semi-infinite plane, discs, et cetera, can be of interest. In such cases, the focus is on the effects of geometry and confinement on the plasmon energy.   
Such studies rely on solutions of the Maxwell equations whose 
main input is the conductivity or dielectric function of the material in the long wave-length limit, i.e. for $q \to 0$.
Our theory can be incorporated into such numerical schemes, which require the local conductivity as their input. 

To leading order in $q \ll k_F$, we find (expressions for these functions holding at larger values of $q$ can be found in Appendix~\ref{sec:explicit_formula}):
\begin{align}
\sigma_{c}(q,\omega)
&\simeq\frac{\sigma_{D} }{1-i\omega \tau_{\parallel}}\left[1+\frac{2 \gamma^2}{(1-i\omega\tau_{\parallel})^{2}}\right],\\
\epsilon(q,\omega)-1
&\simeq \frac{-\omega_{p}^{2}(q)}{\omega (\omega+ \frac{i}{\tau_{\parallel}})}\left[1+\frac{2 \gamma^2}{(1-i\omega\tau_{\parallel})^{2}}\right]. \label{eq:diel} 
\end{align}
In the last expression, $\omega_{p}(q)=\sqrt{4\alpha_{gr}k_F v_{F}^{2} q}$ and $\sigma_{D}=\frac{\rho_{0}e^2\tau_{\parallel} }{k_F/v_{F}}$ are the plasmon frequency and Drude weight, $\rho_0$ is the carrier density, and $k_F$ ($v_F$) is the Fermi momentum (velocity). Note that the corrections to the conductivity and dielectric function due to skew scattering are second order in $\gamma = \tau_{\parallel}/\tau_{\perp}$. This is because the SHE induces a spin current in the transverse direction with magnitude proportional to  $\gamma$. However, through the inverse SHE, the spin current produces a charge current in the longitudinal direction, counteracting the original charge current, which introduces another factor of $\gamma$. Thus,
even for the large SHE like the one observed in Ref.~\onlinecite{jaya2014giant} for
which $\gamma \sim 0.1$, the correction to the response functions can be small (i.e. few percent).  This  is also the reason why the effect of SOC disorder on the plasmon properties are rather weak. 

 Finally, let us point out that, in the long wavelength limit ($q\to 0$), the frequency dependence of the spin Hall angle is given by
\begin{equation}
\Theta_{\mathrm{SH}}(q,\omega)
\simeq \frac{\gamma}{1-i\omega\tau_{\parallel}},
\label{eq:thetash}
\end{equation}
which reduces to $\Theta_{\mathrm{SH}} = \gamma$ in the DC limit. 
It is interesting that the DC spin Hall angle is independent
of the order in which the $q \to 0$ and $\omega \to 0$
limits is taken. This is quite different from the behavior 
exhibited by the conductivity and the dielectric function, whose non-analyticity yields a different result depending
on whether $q \to 0$  is taken before or after $\omega \to 0$. The origin of this non-analyticity is the existence of particle-hole excitations in the spectrum
of the metal, which is a consequence of the Fermi statistics.
On the other hand, the analytical behavior exhibited by the spin Hall angle $\Theta_{\mathrm{SH}}(q,\omega)$ is akin to the behavior of the Hall angle in the  Drude theory of the classical Hall effect, which for $q\to 0$  takes a similar form  to Eq.~\eqref{eq:thetash} with $\gamma$ replaced by 
$\omega_c\tau_{\parallel}$, $\omega_c$ being the cyclotron frequency. The analogy implies that the 
the behavior of the spin Hall angle is unaffected 
by Fermi statistics, that is, it emerges entirely from skew 
scattering, which is a single-particle effect.

\section{Boltzmann transport equation}
\label{sec:BTE}

This section describes the semi-classical Boltzmann Transport Equation (BTE) used to derive the results presented in 
Sec.~\ref{sec:results}.  We consider a graphene sheet decorated with a dilute ensemble of randomly distributed impurities that induce SOC by proximity, such as adatom clusters.~\cite{jaya2014giant,balakrishnan2013colossal}
In other words, the graphene electrons are subjected to a random SOC potential.

We assume the graphene is electron-doped (hole-doped graphene will exhibit similar physics, due to the particle-hole symmetry of the Hamiltonian of pristine graphene).  The transport properties can be calculated using the semi-classical BTE,
\begin{align} \label{eq:BTE}
\partial_{t}n_{\vec{p}\sigma}(\vec{r},t)&
+\vec{v}_{\vec{p}}  \cdot \vec{\nabla}_{\vec{r}} n_{\vec{p}\sigma}(\vec{r},t) 
\nonumber \\
&+ e\vec{E}(\vec{r},t) \cdot \vec{\nabla}_{\vec{p}} n_{\vec{p}\sigma}(\vec{r},t)  = I[n_{\vec{p}\sigma}(\vec{r},t)],
\end{align}
where $n_{\vec{p}\sigma}(\vec{r},t)$ is the electron distribution function for momentum $\vec{p}$ and spin $\sigma$.  The spin quantization axis is perpendicular to the graphene plane, which we take to be the $z$ axis, and $\sigma=\pm 1$ stands for spin pointing in the $\pm\vec{\hat{z}}$ direction. Henceforth, we use the short-hand notation $\bar{\sigma} = -\sigma$, $\vec{v}_{\vec{p}} = v_{F} \vec{\hat{p}}$ ($\vec{\hat{p}} = \vec{p}/|\vec{p}|$) is the group velocity, $v_F$ is the Fermi velocity, and $e < 0$ is the electron electric charge; $\vec{E}(r,t)$ is the total electric field:
\begin{equation} \label{eq:Efield}
\vec{E}(\vec{r},t)= \vec{E}_{ext}(\vec{r},t) - \vec{\nabla}_{\vec{r}} 
 	\int d^{2}\vec{r}^{\prime} \frac{\rho(\vec{r}',t)}{ | \vec{r} -\mathbf {r'} |}.
\end{equation}
Here, $\vec{E}_{ext}(\vec{r},t)=E_{ext}(\vec{q},\omega)e^{i(\vec{q} \cdot \vec{r} -\omega t)}\, \vec{\hat{x}}$ is the external applied electric field, taken to point in the $\vec{\hat{x}}$ direction, and $\rho(\vec{r},t)=e \sum_{\vec{p},\sigma} n_{\vec{p},\sigma}(\vec{r},t)$ is the charge density.

The BTE applies to the long wavelength limit where   $q\ll k_F$ and $\omega \ll \epsilon_{F}$, $k_F$ ($\epsilon_{F}$) being the Fermi momentum (energy) of doped graphene. The electric field $\vec{E}(\vec{r},t)$ in Eq.~\eqref{eq:Efield} must be determined self-consistently, akin to the quantum mechanical treatment of the density-density response function within the random phase approximation (RPA).  We have neglected corrections to the BTE arising from the interactions beyond the (time-dependent) Hartree potential since we are interested in the long wave-length properties only, for which the RPA is most accurate.

Next, we assume that the dominant momentum relaxation mechanism arises  from scattering with impurities, which is a good approximation at temperatures $T \ll \epsilon_F/k_B$. In the \emph{dilute} impurity limit, the collision integral $I[n_{\vec{p}\sigma}]$ takes the following 
form:~\cite{kohn_luttinger_2}
\begin{equation}
\label{eq:coll_int}
I[n_{\vec{p}\sigma}]=\sum_{\sigma}\int\frac{d^{2}k}{(2\pi)^{2}}
\left( n_{\vec{k}\sigma}
- n_{\vec{p}\sigma}\right)W_{\sigma}(\vec{k} \rightarrow \vec{p} ),
\end{equation}
where
\begin{equation}
\label{eq:W}
W_{\sigma}(\vec{k}\rightarrow \vec{p} )= 2\pi n_{i} \delta(\epsilon_{p}-\epsilon_{k}) \left|T_{\sigma}(\vec{k} \rightarrow \vec{p} )\right|^{2}
\end{equation}
is the total rate of scattering by the impurities, $n_{i}$ is the mean impurity (areal) density, and $T_{\sigma}(\vec{k} \rightarrow \vec{p} )$ is the single-impurity $T$-matrix projected onto the band of the carriers at the Fermi level. For the sake of simplicity and to ascertain the effects of skew-scattering only, in this study we consider SOC disorder that conserves the spin projection along the $z$-axis (which is taken perpendicular to the graphene plane).  This requires $[T, s_{z}]=0$, where $s_{z}$ is spin 
$z$-component. In other words, the proximity-induced SOC disorder must be of the intrinsic (Kane-Mele) type,~\cite{Ferreira2014}  which is sufficient to explain the data for the DC charge and spin-Hall conductivity from the experiment of 
Ref.~\onlinecite{jaya2014giant} (see 
Appendix~\ref{app:micro_model} for details).

\subsection{Spin-dependent drift velocity ansatz}

We solve the BTE by using  the following \emph{ansatz}:
\begin{equation}
n_{\vec{p}\sigma}(\vec{r},t) = \frac{1}{\exp(\beta ( \epsilon_{p} -\mu - \vec{v}_{\sigma}(\vec{r},t) \cdot \vec{p} ))+1}.
\end{equation}
Here, $\beta = 1/k_B T$ is the inverse absolute temperature ($k_B$ is Boltzmann's constant), $\vec{v}_{\sigma}(\vec{r},t)$ is an undetermined drift velocity which depends on the spin $\sigma$  (we assume that $v_{\sigma}(\vec{r},t) \propto E$); $\mu$ is the global chemical potential of the system, and $\epsilon_{p}=v_{F} p$ is the energy of the graphene electron. To linear order in $\vec{v}_{\sigma}(\vec{r},t)$, the distribution function can be expanded as
\begin{align}
n_{\vec{p}\sigma}(\vec{r},t) &= n^{0}(\epsilon_{p})+ \delta n_{\vec{p}\sigma}(\vec{r},t), \\
\delta n_{\vec{p}\sigma}(\vec{r},t) &\equiv \left(-\frac{\partial n^{0}(\epsilon_{p} ) }{\partial \epsilon}\right)\vec{v}_{\sigma} (\vec{r},t) \cdot \vec{p}. \label{eq:ansatz}
\end{align}
Here, $n^{0}(\epsilon_{p})$ is the equilibrium Fermi-Dirac distribution:
\begin{equation}
n^{0}(\epsilon_{p})=\frac{1}{e^{\beta(\epsilon_{p} -\mu)}+1}.
\end{equation}
The charge current $\vec{J}_{c} =e g_{v}\sum_{\vec{p}\sigma} \vec{v}_{p} n_{\vec{p}\sigma}$, and the spin current $\vec{J}_{s} = e g_{v}\sum_{\vec{p}\sigma}\sigma\vec{v}_{p}n_{\vec{p}\sigma}$, are given by
\begin{align}\label{eq:charge current def}
\vec{J}_{c}(\vec{r},t) &= \frac{e\rho_{0}}{2}\left( \vec{v}_{+}(\vec{r},t) + \vec{v}_{-}(\vec{r},t)\right), \\
\label{eq:spin current def}
\vec{J}_{s}(\vec{r},t) &= \frac{e\rho_{0}}{2}\left( \vec{v}_{+}(\vec{r},t)-\vec{v}_{-}(\vec{r},t) \right),
\end{align}
where $ \rho_{0}=\sum_{\vec{p}}n^{0}(\epsilon_{p}) =  g_{v} g_s k_F^{2}/4\pi$ is the total electron density and $g_s = g_v = 2$ are the spin and valley degeneracies. Note that the spin current is measured in the same units as the 
charge current, which means that it does not contain the extra factor of $\frac{1}{2}$ (in $\hbar = 1$ units) arising from the electron spin angular momentum.

Using the above  \emph{ansatz} and the general form of single-impurity $T$-matrix,  the explicit form of the collision integral is obtained (the details are given in Appendix~\ref{app:lifetime}):
\begin{align}\label{eq:coll-int}
I[\delta n_{\bf{p},\sigma}] &= - \frac{1}{\tau_{\parallel}(\mu)}
\delta n_{\vec{p}\sigma} (\vec{r},t)+ \left(\frac{\partial n^{0} }{\partial\vec{p}}\right)\cdot \vec{F}_{\sigma}(\vec{r},t), \\
\label{eq:mag_force}
\vec{F}_{\sigma} (\vec{r},t) &\equiv \frac{\epsilon_{F}}{\tau_{\perp}(\mu)v_{F}^{2}}(\vec{v}_{\sigma}(\vec{r},t)\times\vec{\hat{z}})\sigma.
\end{align}
Expressions for the elastic mean free time, $\tau_{\parallel}(\mu)$, and the skew scattering time, $\tau_{\perp}(\mu)$, in terms of the $T$-matrix elements are given in 
Appendix~\ref{app:lifetime}.  They depend on the chemical potential $\mu$. However, in order to lighten the notation,  we will suppress the dependence  on $\mu$ in what follows. In Eq.~\eqref{eq:mag_force}, $\vec{F}_{\sigma}(\vec{r},t)$ is an effective Lorentz force which arises from the skew scattering of the SOC disorder-potential. Similar to how the Lorentz force drives a transverse charge current in the Hall effect, this effective ``spin-Lorentz force'' $\vec{F}_{\sigma}$ drives a transverse spin current. Thus, in the absence of SOC disorder where $\tau^{-1}_{\perp} \rightarrow 0$,  $\vec{F}_{\sigma}$ vanishes.  
The magnitude of the spin-Lorentz force depends on the Fermi energy and $\tau_{\perp}$.  For spin up (down) electrons, $\vec{F}_\sigma$ points perpendicular to the spin drift velocity of the electron and the positive (negative) spin quantization axis $\vec{\hat{z}}$.

%
\subsection{Particle-number conserving approximation} \label{secr:rta_local_equiv}

The conservation of electric charge (i.e.~the continuity equation) follows by summing the BTE, Eq.~\eqref{eq:BTE}, over  momentum $\vec{p}$ and spin $\sigma$.~\cite{landau_lifshitz_10} This places a constraint on the collision integral: $\sum_{\vec{p},\sigma}I[\delta n_{\vec{p},\sigma}]=0$. However, an ansatz for $\delta n_{\vec{p},\sigma}$ may fail to fulfill the particle-number conservation constraint (see Ref.~\onlinecite{kragler1980dielectric} and references therein). Following Mermin's prescription\cite{MerminPhysRevB.1.2362} to satisfy the conservation law,  the collision integral must be modified to ensure that the distribution function relaxes to the local (rather than the global, $n^{0}(\epsilon_{p})$) equilibrium distribution function,~\cite{kragler1980dielectric}
\begin{align}
g_{p}(\vec{r},t) &= \frac{1}{e^{\beta(\epsilon_{p}-\mu-\delta\mu(\vec{r},t))}+1} \nonumber \\
&\approx n^{0}(\epsilon_{p})-\frac{\partial n^{0}(\epsilon_{p}) }{\partial\epsilon_{p}}\delta\mu(\vec{r},t),
\end{align}
where $\delta\mu(\vec{r},t)$ is the local chemical potential. Thus, the modified collision integral reads
\begin{multline}\label{eq:coll-int-mermin}
\tilde{I} [ n_{\bf{p},\sigma}] = - \frac{1}{\tau_{\parallel}}
\left[n_{\vec{p}\sigma}(\vec{r},t) - g_{p}(\vec{r},t) \right]
\\+ \left(\frac{\partial n^{0}(\epsilon_{p})}{\partial\vec{p}}\right)\cdot \vec{F}_{\sigma}(\vec{r},t).
\end{multline}
The local chemical potential, $\delta\mu(\vec{r},t)$, is determined from the constraint $\sum_{p} I[ n_{\vec{p}\sigma}]=0$, which yields:
\begin{equation}
\delta\mu(\vec{r},t)=\frac{1}{ \sum_{\vec{p}} \frac{\partial n^0}{\partial \epsilon}} \sum_{\vec{p} }\delta n_{\vec{p},\sigma}(\vec{r},t).
\end{equation}
The modified collision integral, Eq.~\eqref{eq:coll-int-mermin} leads to response functions that interpolate  between the collisionless ($ \omega \tau_{\parallel} \gg 1$) and the hydrodynamic regimes ($\omega \tau_{\parallel} \ll 1$).~\cite{giuliani_giovanni} In this regard, in Appendix \ref{app:hydrodynamic}, we show that, as the system enters the hydrodynamic regime, the results of the present particle-number conserving approach to the results of a harmonic mode ansatz, which is appropriate in the  hydrodynamic regime.

\subsection{Generalized Ohm's law} 
\label{sec:generalized_ac_formula}

We are now ready to solve the linearized BTE and derive the generalized Ohm's law in the presence of SOC.  Recall that the self-consistent electric field has the form
\begin{equation}
  \vec{E}(\vec{r},t)=E(\vec{q},\omega)e^{i(\vec{q} \cdot \vec{r} -\omega t)}\, \vec{\hat{x}}. 
\end{equation}
Without loss of generality, we focus on the relevant Fourier component of $n_{\vec{p}\sigma}(\vec{r},t)
= n_{\vec{p}\sigma}(\vec{q},\omega)e^{i(\vec{q}\cdot\vec{r}-\omega t)}$. For notational simplicity, we henceforth suppress the $\vec{q}$ and $\omega$ dependence on $\delta n_{\vec{p},\sigma}$, $v_{\sigma}$, and $E$. Expanding Eq.~\eqref{eq:BTE} to linear order in $E$ now gives
\begin{equation}
\label{eq:lin_boltzman}
i( \vec{v}_{\vec{p}} \cdot \vec{q} -  \omega)  \delta n_{\vec{p}\sigma} 
+\frac{dn^{0}(\epsilon_{p})}{d\epsilon} \vec{v}_{\vec{p}} \cdot   e\vec{E}
 = \tilde{I} [\delta n_{\vec{p}\sigma}],
\end{equation}
with the local particle conserving collision 
integral~\cite{MerminPhysRevB.1.2362,kragler1980dielectric,giuliani_giovanni}
\begin{multline}
\tilde{I} [ \delta n_{\bf{p},\sigma}] = - \frac{1}{\tau_{\parallel}}
\left( \delta n_{\vec{p}\sigma} -  \frac{1}{N(\epsilon_{F})} \sum_{p}\delta n_{\vec{p},\sigma}\right)  \\
+ \left(\frac{\partial n^{0}(\epsilon_{p})}{\partial\vec{p}}\right)\cdot \vec{F}_{\sigma}
\end{multline}
Substituting the \textit{ansatz} \eqref{eq:ansatz} into Eq.~\eqref{eq:lin_boltzman} leads to equations~\eqref{eq:result1} and \eqref{eq:result2}.
These equations are the generalized Ohm's law whose consequences have been discussed in Sec.~\ref{sec:results}. The ``skew scattering resistivity'' $\rho_{ss}= \frac{k_F/v_{F} }{ \tau_{\perp}e^{2} v_{F} \rho_{0}}$ has the same form as the Drude resistivity in graphene but with the elastic mean free time $\tau_{\parallel}$ replaced by the skew scattering time $\tau_{\perp}$, and $\rho_{0}=g_{v}k_F^{2}/(2\pi)$ is the density of electrons in graphene. Multiplying  the skew-scattering resistivity $\rho_{ss}$ by the Drude conductivity yields the spin-dependent electric field, which drives the transverse spin current.

In Eq.~\eqref{eq:result1}, $\sigma_{m}(q,\omega)$ represents the local particle conserving charge conductivity, given by
\begin{align} 
\sigma_{m}\left(q,\omega \right)
&=\frac{ \sigma_{0} \left(q,\omega \right) }
{ M\left(q,\omega\right)}, \\
M\left(q,\omega\right) &=\frac{1}{N(\mu)}\sum_{\vec{p}}\left(-\frac{\partial n^{0}}{\partial\epsilon}\right)\frac{\omega-\vec{v_{p}}\cdot\vec{q}}{\omega-\vec{v_{p}}\cdot\vec{q}+\frac{i}{\tau_{\parallel}}}, \label{eq:mermin_factor}\\
\sigma_{0} \left(q,\omega \right) &= \frac{4 ie^{2}\omega}{q^2}
 \sum_{\vec{p}}\left(-\frac{\partial n^{0}}{\partial\epsilon}\right)\frac{\vec{v_{p}}\cdot\vec{q}}{\omega-\vec{v_{p}}\cdot\vec{q}+\frac{i}{\tau_{\parallel}}}.
\end{align}
Here $M(q,\omega)$ is the dimensionless correction arising from the particle-number conserving collision integral introduced in Sec.~\ref{secr:rta_local_equiv}. It corrects $\sigma_{0}(q,\omega)$, which does not conserve local particle number, so as to produce a conductivity that is accurate both in hydrodynamic and collisionless regimes. Using the charge continuity equation, we can also define the corrected  Lindhard function:
\begin{equation}\label{eq:chi_sigmac}
\chi_{m}\left(q,\omega \right)= 
\frac{q^2}{ie^2 \omega} \sigma_{m}\left(q,\omega \right).
\end{equation}
Note that $\chi_{m}\left(q,\omega \right)$  corresponds to the semi-classical limit of the Lindhard function in the presence of scalar impurities, and can be derived from Mermin's result~\cite{MerminPhysRevB.1.2362} (obtained using the quantum Boltzmann equation) in the limit where $q \ll k_F$ and $\omega \ll \epsilon_F$.

Finally, the spin Hall angle $\Theta_{\mathrm{SH}}$ is defined as the ratio of the spin current to charge current:
\begin{equation} \label{eq:spin_hall_angle}
\Theta_{\mathrm{SH}} (q,\omega)
=\left(\frac{i \epsilon_{F}}{\tau_{\perp}\rho_{0}v_{F}^{2}}\right)\sum_{\vec{p}}\left(-\frac{\partial n^{0}}{\partial\epsilon}\right) 	\frac{ \left( \vec{E}/E	\cdot (\vec{v}_{p}\times \vec{\hat{z}}) \right)^2 } {\omega-\vec{v_{p}}\cdot\vec{q}+\frac{i}{\tau_{\parallel}}}.
\end{equation}
The spin Hall angle stems from the spin Lorentz force introduced in Eq.~\eqref{eq:mag_force}.  Thus, we can interpret the charge current in Eq.~\eqref{eq:result1} as being driven by the combination of electric field and a field produced by the spin current. 

The expressions for the response functions derived in this section from the solution of the BTE are  explicitly evaluated in Appendix~\ref{sec:explicit_formula}.

\section{Summary and Outlook}\label{sec:summ}

In this article, we have theoretically investigated the impact of spin-orbit coupling (SOC) disorder on the electrodynamics of two dimensional electron (hole) gases, focusing on doped graphene. This has been achieved by developing a formalism  that generalizes previous treatments of the semiclassical transport equations to account for the response of the system to non-uniform time-dependent electric fields. The formalism has allowed us to obtain the explicit forms of the non-local frequency dependent conductivity, dielectric function, and spin Hall angle in the presence of skew scattering. The latter mechanism is responsible for the extrinsic spin Hall effect (SHE), by which a longitudinal charge current is converted into a transverse spin current. 

 In addition, we have applied our formalism  to  analyze the effects of SOC disorder on the plasmon dispersion. We have thus found that SOC disorder also decreases the plasmon lifetime and leads to a softening (i.e. red shift relative to the clean system plasma frequency) of the plasmon dispersion. The softening caused by skew scattering is different from the one found for scalar disorder.~\cite{quinn_giuliani} However, as the wave number of the 
plasmon increases, the corrections to the dispersion arising from skew scattering become smaller to 
the extent that the it may be hard to discern their effect in a real experiment. This finding suggests that the plasmon softening and linewidth are largely unaffected the SOC disorder that causes the extrinsic SHE.
The reason is that, even if a faction $\gamma$ of the longitudinal electric current associated with the plasmon is converted into a transverse spin current by the SHE, the electromagnetic effect of the latter is proportional to $\gamma^2$, i.e. the square of the DC spin Hall angle.

It is interesting to discuss the main differences of the electrodynamics developed here with electrodynamics on the surface of 3D topological insulators.  Our theory obtains an oscillating transverse spin current is generated when graphene decorated with adatoms is subjected to an AC electric field. This is quite unlike the AC spin currents observed on the surface of 3D topological insulators~\cite{di2013observation}, which are longitudinal due to the spin-momentum locking taking place at the surface of those materials. In graphenne,  this is because the spin current is induced by the SOC disorder via the skew scattering mechanism, which leads to an effective spin-dependent Lorentz force as discussed in Sec.~\ref{sec:BTE}.

Future  extensions of this work (currently underway \cite{in_prep}) include accounting for spin-flip mechanism by the SOC disorder, and extending the present formalism beyond the semiclassical regime to include, e.g. the effects of inter-band transitions. Concerning the former, preliminary results\cite{in_prep} indicate that
the picture put forward here is not substantially modified by the presence of other forms of SOC that induce spin-flip scattering. 

Finally, we hope that this work will spur the experimental interest to search for SOC-related effects in plasmonics. In the case of graphene, we believe this may be possible through an accurate measurement and comparison of the electrodynamic response of CVD and exfoliated graphene. Unfortunately, the results of our study seem to indicate that  from the plasmon properties alone, such effects will be hard to infer. However,  other probes may be devised to experimentally detect the alternating spin currents created by the AC SHE.  We believe such studies will shed additional light on the mechanisms driving the existence of non-local DC 
currents.

\acknowledgements 

MAC work is supported by the Ministry of Science and Technology (Taiwan)
and Taiwan's National Center of Theoretical Sciences (NCTS). MAC and CH
thank the hospitality of the Donostia International Physics Center (DIPC),
in San Sebastian (Spain), where a part of this research was carried out. CH and YC were supported in part by the Singapore National Research
Foundation grant No.~NRFF2012-02.

\appendix

\section{Microscopic model of SOC disorder} \label{sec:single_impurity}
\label{app:micro_model}

In this appendix, we provide the details of the single-impurity $T$-matrix used in the calculation of the collision integral. The form of $T$ is constrained by the symmetries of the total Hamiltonian $H=H_{0}+w(\vec{r})$ where $H_{0}$ is the Hamiltonian of pristine graphene and $w(\vec{r})$ is the (single) impurity potential created by one absorbate (e.g. a cluster of adatoms). In the continuum limit, we model pristine graphene using the $\vec{k}\cdot\vec{p}$ Hamiltonian:
\begin{equation}
  H_{0}= v_{F} \left( \tau_{z}\sigma_{x}p_{x}+\sigma_{y}p_{y} \right),  
\end{equation}
where $v_{F}$ is the Fermi velocity, and $\sigma_{\alpha}$, $\tau_{\alpha}$, $s_{\alpha}$ ($\alpha = x,y,z$)  are the Pauli matrices in the Hilbert spaces of the sublattice, valley and electron spins, respectively. Thus, in the continuum limit, the $T$-matrix would be in general a linear combination of 16 matrices $s_{\alpha}\tau_{\beta}$ ($\alpha,\beta=0$ corresponds to the unit matrix).

However, for clusters of adatoms with characteristic size $R \gg a$, where $a = 2.46\,$\AA\, is the interatomic distance in graphene, inter-valley scattering is 
suppressed.~\cite{Ferreira2014,hector2012spin-orbit} Neglecting inter-valley scattering, the $T$-matrix contains terms proportional to $\tau_{0}$ and $\tau_{z}$. Further assuming that the impurity potential is time-reversal invariant and rotationally invariant, the fully 
symmetry-constrained  form of the the $T$-matrix is obtained:~\cite{Ferreira2014}
\begin{equation}
\langle \vec{p}|T| \vec{k}\rangle = a\; s_{0}\tau_{0} + (b \;s_{z} + c \;(\vec{p}-\vec{k})\cdot \vec{s})(\vec{\hat{k}}\wedge\vec{\hat{p}})
\end{equation}
where $\hat{\vec{k}} = \vec{k}/k$, $\hat{\vec{p}} = \vec{p}/p$, and $\vec{\hat{k}}\wedge\vec{\hat{p}}=\sin \theta$, $\theta$ being the scattering angle,  and $\vec{s}=(s_{x},s_{y})$. Here, $a,b,c$ are functions of $\vec{\hat{k}} \cdot \vec{\hat{p}}=\cos\theta$ which depend on the microscopic details of the scatterer potential. Note that $c$  corresponds to the probability amplitude for spin-flip scattering, which we shall neglect because our study only focuses on the effects of the skew scattering 
that is described by the term proportional to $b$.
This type of models was also a minimal model able to reproduce the experimental data for the DC spin Hall angle and the charge conductivity in Ref.~\onlinecite{jaya2014giant}.

 As an example, in order to illustrate a particular form of the $T$-matrix resulting from a concrete microscopic model,  let us consider the following Dirac-delta potential:
\begin{align} \label{eq:wr_sc}
  w(\vec{r}) =  ( \lambda_0 + \lambda_{i} \sigma_z \tau_z s_z )\delta(\vec{r}).
\end{align}
The first term in Eq.~\eqref{eq:wr_sc} corresponds to a scalar (spin-independent) potential and the second term corresponds to SOC of the intrinsic (or  Kane-Mele~\cite{kane2005quantum}) type. Notice that both terms in $w(\vec{r})$ potentials commute with $s_{z}$. Thus, within this model, there is no  spin flip scattering.  
Here $\lambda_{0}$ and $\lambda_{i}$ parametrize  the scalar and SOC disorder potential strength. Following Ref~\onlinecite{hungyu2015}, the $T$-matrix of this model is computed as,
\begin{equation}\label{eq:t_sc}
\langle \vec{p}|T| \vec{k}\rangle = \gamma_{0}\cos \left( {\frac{ \theta }{2}} \right) s_{0}+
i \gamma_{i} \sin \left( {\frac{\theta}{2}}  \right) s_{z}.
\end{equation}
$\vec{\hat{k}} \cdot \vec{\hat{p}}=\cos\theta$ is the scattering angle between the incident momentum $\vec{p}$ and scattered electron $\vec{k}$. Note that the spin of the incident and scattered electron is not specified, so the $T$-matrix is a matrix in spin space. The couplings in the above
equation are given by:
\begin{equation}
\gamma_{0}=\frac{\lambda_{0} + G_{0}(k)(\lambda_{0}^{2} -\lambda_{i}^2 ) } {(G_{0}(k)\lambda_{0}-1)^2- (G_{0}(k)\lambda_{I})^2},
\end{equation}
\begin{equation}
\gamma_{i}=\frac{\lambda_{i} } {(G_{0}(k)\lambda_{0}-1)^2- (G_{0}(k)\lambda_{i})^2},
\end{equation}
where 
\begin{equation}
G_{0}(k)=\frac{k}{2\pi v_{F}}\log |kR| - i\frac{k}{4v_{F}},
\end{equation}
is the Green function at the origin. Note a short distance cut-off of the order of the inverse of the scatterer radius, $\approx R^{-1}$, has been used to regulate the otherwise divergent integral over $k$. We refer the interested reader to Ref.~\onlinecite{hungyu2015} for details about the derivation of the above results.

\section{Collision integral and scattering rates}
\label{app:lifetime}
In this appendix, the elastic mean free time and skew-scattering time are derived from the spin-conserving $T$-matrix \eqref{eq:t_sc}, by making use of the \emph{ansatz} introduced in Sec.~\ref{sec:BTE}. The matrix elements of the $T$-matrix are give by
\begin{align}
\langle \vec{p}\sigma|T|\vec{k}\sigma'\rangle  =
\left( \gamma_{0} \cos\theta/2 + i\gamma_{i} \sigma \sin \theta/2 \right) \delta_{\sigma,\sigma'}.
\end{align}
By substituting this and Eq.~\eqref{eq:ansatz} into Eq.~\eqref{eq:coll_int}, we can derive the scattering rates. For example, the spin Lorentz force is derived as 
follows (recall that $\vec{\hat{k}} \cdot \vec{\hat{p}}=\cos\theta$):
\begin{align}
I_2 \equiv 2 \pi n_{i} \Big[ \int &  \frac{d^{2}k}{(2\pi)^2} \text{Im}(\gamma_0 \gamma_i)\: \sin \theta \: \delta(\epsilon_k -\epsilon_p)\sigma \nonumber \\
&\left(\vec{\hat{k}}-\vec{\hat{p}} \right)\cdot \vec{v}_{\sigma}(\vec{r},t) \left( -\frac{\partial n^{0}}{\partial \epsilon} \right) k_F \Big].
\end{align}
Since energy is conserved in collision, we can set $k=|\vec{k}|=|\vec{p}|$ ($\vec{\hat{k}} = \vec{k}/k$, $\vec{\hat{p}} = \vec{p}/k$). Upon expanding $\vec{\hat{k}}= \cos\theta \vec{\hat{p}}+\sin \theta (\vec{\hat{z}}\times\vec{\hat{p}})$, we find that only the term proportional to $\sin \theta$ is non-vanishing upon integration. This leaves
\begin{align}
I_2 = 2 \pi n_{i} \Big( \int &  \frac{d^{2}k}{(2\pi)^2} \text{Im(}\gamma_0 \gamma_i)\sin^{2} \theta \delta(\epsilon_k -\epsilon_p)\Big) \nonumber \\
&\sigma \left(\vec{\hat{z}} \times \vec{\hat{p}} \right) \vec{v}_{\sigma}(\vec{r},t) \left( -\frac{\partial n^{0}}{\partial \epsilon} \right) k_F.
\end{align}
After some algebra, we arrive at the scattering rate for the (effective) Lorentz spin Hall force:
\begin{equation}
\tau^{-1}_{\perp}=- 2 \pi n_{i} \int \frac{d^{2}k}{(2\pi)^2} \text{Im(}\gamma_0 \gamma_i)\sin^{2} \theta \:  \delta(\epsilon_k -\epsilon_p).
\end{equation}
As mentioned in Appendix~\ref{sec:single_impurity}, the $T$-matrix couplings $\gamma_0$ and $\gamma_i$ depend on the incoming (electron) energy $\epsilon_k = v_{F}k$ and the strength of the  scalar and SOC potential $\lambda_0$, and $\lambda_i$. The sign of the skew scattering rate can be either positive or negative. However, the inverse of the elastic mean free time is always positive and given by
\begin{align}
\tau^{-1}_{\parallel}=2 \pi n_{i} \int \frac{d^{2}k}{(2\pi)^2} &\left(|\gamma_{0}|^{2} \cos^{2}(\theta/2) +|\gamma_i|^{2}\sin^{2}(\theta/2) \right) \nonumber \\ &(1-\cos\theta)\delta(\epsilon_{p}-\epsilon_{k}).
\end{align}
%

\section{Comparison between Mermin's approach and Harmonic Mode Ansatz}
\label{app:hydrodynamic}
\begin{figure*}[ht]
\includegraphics[width=0.95\textwidth]{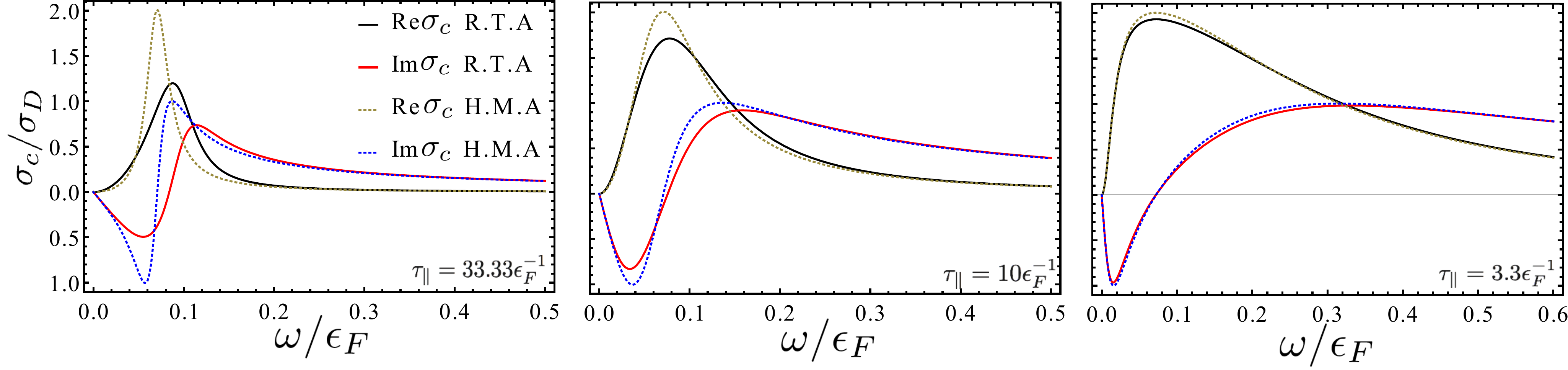}
\caption{
Charge conductivity calculated within the particle-number conserving relaxation time approximation (RTA) and a harmonic mode ansatz (HMA) which is exact in the hydrodynamic regime is  plotted against frequency. In order to display the differences as the system is tuned into the hydrodynamic regime, we have fixed $q = 0.1 \: k_F$ and varied the mean free time $\tau_{\parallel}$. In the hydrodynamic limit where
 $\tau^{-1}_{\parallel} \gg v_{F}q $ and $\tau^{-1}_{\parallel} \gg \omega $ the results of RTA approach very closely the results of HMA. In the opposite limit, i.e. the 
collisionless limit, the conductivity obtained from the RTA exhibit a broader peak because of the diffusion pole in the conductivity.~\cite{giuliani_giovanni}} \label{fig:hma_rta}
\end{figure*}
In this appendix, we compare the response functions obtained using the Boltzmann transport equation (BTE) formulated within Mermin's particle-number conserving relaxation time approximation (RTA) and  a harmonic mode 
ansatz  (HMA) that describes in the  hydrodynamic regime where $v_F q,\omega \ll \tau^{-1}_{\parallel}$. In the latter regime, equilibration  takes place very fast and we can use the following form of the BTE,
\begin{equation} \label{eq:lin_boltz}
i( \vec{v}_{\vec{p}} \cdot \vec{q} -  \omega)  \delta n_{\vec{p}\sigma} 
+\frac{dn^{0}(\epsilon_{p})}{d\epsilon} \vec{v}_{\vec{p}} \cdot   e\vec{E}
 = I [\delta n_{\vec{p}\sigma}],
\end{equation}
where the collision integral, $I [\delta n_{\vec{p}\sigma}]$, is given by Eq.~\eqref{eq:coll-int}. In addition, we can make the following ansatz:
\begin{equation}\label{eq:HMA}
\delta n_{\vec{p}\sigma} \approx -\frac{dn_{F}(\epsilon_{p})}{d\epsilon_{p}}
\left( C_{\sigma}+A_{\sigma}\cos\theta_{p}+B_{\sigma}\sin\theta_{p} \right).
\end{equation}
This is justified since higher angular momentum deformations relax very fast in the hydrodynamic regime. Note that this ansatz is the simplest generalization of the one previously used to account for the DC SHE,~\cite{Ferreira2014,hungyu2015} which allows for a change in volume of the Fermi surface (a breathing mode). In the above expression, $\cos \theta_{p} = \vec{\hat{p}}\cdot \vec{\hat{x}}$ denotes the cosine of the angle subtended by $\vec{p}$ with the electric field direction, $\vec{\hat{x}}$. The $A_\sigma$ and $B_\sigma$ terms describe the longitudinal and transverse response, respectively, and  $C_{\sigma}$ describes the breathing mode, which is necessary to describe the plasmon mode. In terms of this \textit{ansatz}, the longitudinal charge conductivity and spin conductivity are given by
\begin{align}
  J_\parallel^c &= e g_{v} \sum_{\vec{p}\sigma} \delta n_{ \vec{p} \sigma} \vec{v_{p}}\cdot \vec{\hat{x}}
  = e v_F N(\epsilon_{F})  A_{c} / 2, \label{eq:hma_result1} \\
J_{\perp}^{s} &=eg_{v}\sum_{p\sigma}  \sigma \delta n_{p\sigma}\vec{v_{p}}\cdot\vec{\hat{y}} 
=e v_{f}N(\epsilon_{F})B_{s}/2. \label{eq:hma_result2}
\end{align}
Here, $N(\epsilon_{F}) = g_{v}\epsilon_{F}/ \pi v_{F}^2$ is the density of states at the Fermi energy and $g_{v}=2$ is the valley degeneracy.

Next, we substitute Eq.~\eqref{eq:HMA} into Eq.~\eqref{eq:lin_boltz}.  Carrying out the angular average, we derive six equations for the six unknowns $\{A_\sigma, B_\sigma, C_{\sigma}\}$, which can be written as:
\begin{align}
  &\begin{bmatrix} \omega + \frac{i}{\tau_\parallel} & -v_Fq & \frac{i}{\tau_\perp} \\
    -v_F q & 2\omega  & 0 \\
    -\frac{i}{\tau_\perp} & 0 & \omega + \frac{i}{\tau_\parallel}
  \end{bmatrix}
  \begin{bmatrix}
    A_s \\ C_s \\ B_c
  \end{bmatrix}
  = 0, \label{homogenous} \\
  &\begin{bmatrix}
    \omega + \frac{i}{\tau_\parallel} & -v_F q & \frac{i}{\tau_\perp} \\
    -v_F q & 2\omega & 0 \\
    -\frac{i}{\tau_\perp} & 0 & \omega + \frac{i}{\tau_\parallel}
  \end{bmatrix}
  \begin{bmatrix}
    A_c \\ C_c \\ B_s
  \end{bmatrix} =
  \begin{bmatrix}
    2iev_F E \\ 0 \\ 0
  \end{bmatrix}.
  \label{inhomogenous}
\end{align}
Here, we have re-organized the \textit{ansatz} parameters into spin and charge sectors by defining  $A_s = A_{\uparrow} - A_{\downarrow}$ and $A_c = A_{\uparrow} + A_{\downarrow}$, and similarly definitions for $B_{c},B_s$ and $C_c,C_s$. In addition, the scattering rates $\tau^{-1}_{\parallel}=\Gamma_{\sigma}^{S}$ and $\tau^{-1}_{\perp}=\Gamma_{\sigma}^{0} -\Gamma_{\sigma}^{C}$. The three scattering rates $\Gamma_{\sigma}^{0}$, $\Gamma_{\sigma}^{S}$ and $\Gamma_{\sigma}^{C}$ are defined as follows:
\begin{align}
\frac{\Gamma_{\sigma}^{S} }{N(\mu)} &= \int\frac{ d\theta_{p} d\theta_{k}}{(2\pi)^{2}} W_{\sigma}(\vec{k}\rightarrow \vec{p} ) \sin(\theta_{p}-\theta_{k}), \\
\frac{\Gamma_{\sigma}^{C} }{N(\mu)} &=  \int\frac{ d\theta_{p} d\theta_{k}}{(2\pi)^{2}} W_{\sigma}(\vec{k}\rightarrow \vec{p} ) \cos(\theta_{p}-\theta_{k}), \\
\frac{\Gamma_{\sigma}^{0} }{N(\mu)} &=  \int\frac{d\theta_p d\theta_{k}}{(2\pi)^{2}} W_{\sigma}(\vec{k}\rightarrow \vec{p} ).
\end{align}
The scattering rate $W_{\sigma}(\vec{k}\rightarrow \vec{p})$ is defined in Eq.~\eqref{eq:W}. The ratio $\gamma = \tau_{\parallel}/\tau_{\perp}$ is the (zero temperature) spin Hall angle~\cite{Ferreira2014}, which measures the relative magnitude of transverse spin current over longitudinal charge current. Eq.~\eqref{homogenous} states that the applied AC electric field does not couple to the longitudinal spin response ($A_s$), the net spin response ($C_s$), or the transverse charge response ($B_c$), as might be expected on physical grounds.  We hence ignore these three quantities.  The terms which do couple to the electric field, collected in Eq.~\eqref{inhomogenous}, are the longitudinal charge response ($A_c$) and net charge response ($C_c$), as well as the transverse spin response ($B_s$).  The solution is
\begin{align}
A_{c} &= \frac{2iev_FE}{\omega + i/\tau_{\parallel} - \frac{1}{\tau_{\perp}^{2}}\frac{1}{\omega+ i /\tau_{\parallel}} - \frac{(v_{F}q)^{2}}{2\omega}} \label{eq:Asol} \\
B_{s} &= \frac{i/\tau_{\perp}}{\omega+i/\tau_{\parallel}} \, A_{c} \label{eq:Bsol} \\
C_{c} &= \frac{v_{F}q}{2\omega} \, A_{c}. \label{eq:Csol}
\end{align}
Substituting these solutions into Eq.~\eqref{eq:hma_result1} and Eq.~\eqref{eq:hma_result2} gives us the charge and spin response in the hydrodynamic limit. Thus, as an example, in Fig.~\ref{fig:hma_rta} we show that the agreement of charge conductivity computed within Mermin's RTA and the hydrodynamic ansatz employed in this appendix 
agrees as the hydrodynamic regime (where $\tau^{-1}_{\parallel}\gg \omega$ and $\tau^{-1}_{\parallel} \gg v_{F}q$) is approached.  On the other hand, in the collisionless regime, the results from the harmonic-mode ansatz are inaccurate since the ansatz of Eq.~\ref{eq:HMA} neglects large angular momentum deformations of the Fermi surface.

\section{Evaluation of response functions}
\label{sec:explicit_formula}

In this appendix, the explicit formula of the dielectric function  (cf. Eq.~\ref{eq:dielec}) and charge conductivity  (cf. Eq.~\ref{eq:sigmac}) are provided.  The expressions are most simply written in terms of the dimensionless ratios  $\nu=\omega/(v_{F}q)$ and $\eta=1/(\tau_{\parallel}v_F q)$. In what follows,  we shall use the following formula (see e.g. Ref.~\onlinecite{giuliani_giovanni} for details):
\begin{equation}
I_{1}=\int\frac{d\theta}{2\pi}\frac{1}{\nu+i\eta-\cos\theta}=\frac{1}{\sqrt{(\nu+i\eta)^{2}-1}},
\end{equation}
Hence, the standard Lindhard function (normalized to density of states at the Fermi energy) can be evaluted as follows:
\begin{align}
\frac{\chi_{0}(q,\omega)}{N(\epsilon_{F})} =& \frac{1}{N(\epsilon_{F})}\sum_{\boldsymbol{p}}\left(-\frac{\partial n^{0}}{\partial\epsilon}\right)\frac{\boldsymbol{v_{p}}\cdot\boldsymbol{q}}{\omega-\boldsymbol{v_{p}}\cdot\boldsymbol{q}+\frac{i}{\tau_{\parallel}}}  \nonumber \\
=& \int^{2\pi}_0\frac{d\theta}{2\pi}\frac{v_F q \cos\theta}{\omega+\frac{i}{\tau_{\parallel}}-v_{F}q\cos\theta}
\nonumber \\
=& (\nu+i\eta)I_{1}-1.
\end{align}
Likewise, the "Mermin factor"  
(cf. Eq.~\ref{eq:mermin_factor}) that modified the Lindhard function in order to conserve the particle number is similarly evaluated as follows:
\begin{align}
M\left(q,\omega\right) &=\frac{1}{N(\mu)}\sum_{\vec{p}}\left(-\frac{\partial n^{0}}{\partial\epsilon}\right)\frac{\omega-\vec{v_{p}}\cdot\vec{q}}{\omega-\vec{v_{p}}\cdot\vec{q}+\frac{i}{\tau_{\parallel}}}
\nonumber \\
&= 1-i \eta\:  I_{1}.
\end{align}
Collecting these results, we obtain the  Lindhard function within Mermin's particle-number conserving approximation, 
\begin{align}
 \frac{\chi_{m}(q,\omega)}{N(\epsilon_{F})}&=\frac{(\nu+i\eta)I_{1}-1}{1-i \eta I_{1}}\\
 &=\left(\frac{\nu}{\sqrt{(\nu+i\eta)^{2}-1}-i\eta}-1\right).
 \end{align}
From this expresion, the conductivity can be easily obtained using Eq.~\eqref{eq:chi_sigmac}. Similarly, the dynamic spin Hall angle $\Theta_{\mathrm{SH}}(q,\omega)$ defined in Eq.\eqref{eq:spin_hall_angle} is evaluated as follow
\begin{align}
\Theta_{\mathrm{SH}} (q,\omega)=&\frac{2i}{v_{F}\tau_{\perp}q} \int^{2\pi}_0\frac{d\theta}{2\pi} \frac{\sin^{2}\theta}{\nu+i\gamma-\cos\theta} \nonumber \\
=&\frac{2i}{v_{F}\tau_{\perp}q}
\Bigg[\frac{2}{z_{+} - z_{-}} \nonumber \\
&+\frac{1}{2}\left(\frac{z_{-}^{4}+2z_{-}^{2}+1}{z_{-}^{2}(z_{-}-z_{+})}+\frac{1}{z_{-}}+\frac{1}{z_{+}}\right)\Bigg],
\end{align}
where $z_{\pm}=\nu+i\eta \pm \sqrt{(\nu+i\eta)^2-1}$.   Using the explicit form of $\Theta_{\mathrm{SH}}(q,\omega)$ and $\sigma_{m}(q,\omega)$ the charge conductivity in Eq.\eqref{eq:sigmac} and the dielectric function in Eq.\eqref{eq:dielec} can be obtained. Since they are lengthier than the above expressions and not particularly illuminating, we shall not reproduce them here explicitly.
\bibliographystyle{apsrev4-1}
\bibliography{mybib}
\end{document}